\documentclass[11pt]{article}
\usepackage{amsmath, amssymb, amsfonts, amsthm}

\newcommand{\ph}{\varphi}
\newcommand{\tr}{\mbox{Tr}}
\newcommand{\rd}{{\rm d}}

\newcommand{\bR}{{\mathbb R}}
\newcommand{\bN}{{\mathbb N}}
\newcommand{\cH}{{\cal H}}
\newcommand{\cA}{{\cal A}}
\newcommand{\cJ}{{\cal J}}
\newcommand{\cU}{{\cal U}}
\newcommand{\cL}{{\cal L}}
\newcommand{\cN}{{\cal N}}
\newcommand{\cF}{{\cal F}}
\newcommand{\cE}{{\cal E}}
\newcommand{\cS}{{\cal S}}

\newcommand{\wt}{\widetilde}

\newtheorem{theorem}{Theorem}[section]

\begin{document}

\title{Quantum Dynamics, Coherent States and \\ Bogoliubov Transformations}

\author{B. Schlein \footnote{Partially supported by ERC Starting Grant MAQD-240518} \\ \\ 
Institute for Applied Mathematics, University of Bonn,\\
Bonn, 53115 Germany}

\maketitle

\begin{abstract}
Systems of interest in physics are usually composed by a very large number of interacting particles. At equilibrium, these systems are described by stationary states of the many-body Hamiltonian (at zero temperature, by the ground state). The reaction to perturbations, for example to a change of the external fields, is governed by the time-dependent many-body Schr\"odinger equation. Since it is typically very difficult to extract useful information from the Schr\"odinger equation, one of the main goals of non-equilibrium statistical mechanics is the derivation of effective evolution equations which can be used to predict the macroscopic behavior of the system. In these notes, we are going to consider systems of interacting bosons in the so called Gross-Pitaevskii regime, and we are going to show how coherent states and Bogoliubov transformations can be used to approximate the 
many body dynamics.
\end{abstract}



\section{Introduction}

In these notes, we review some recent results concerning an important question in non-equilibrium statistical mechanics which is the derivation of effective evolution equations approximating the dynamics of many body quantum systems.

When studying systems of interest in physics one typically has to choose between two different approaches. On the one hand, one can opt for a description based on the fundamental laws of physics.
In this case, one would describe the system in terms of its elementary components and their interactions. So a chemical sample would be described as a system of many atoms, interacting through the laws of quantum mechanics. A galaxy could instead be described as a system of many stars interacting through Newtonian or Einstein gravitation.  

One can opt, alternatively, for an effective description of the system, which does not resolve the single components, but instead focus on their collective behavior. For example, the motion of a fluid would then be described by the Euler or the Navier-Stokes equations, a superconductor through the Ginzburg-Landau theory. 

Of course the first approach is more precise, and typically relies on very few assumptions. The second approach, on the other hand, is less precise, it only has a small range of validity, but it is much more accessible to computations and focus exactly on those quantities which are measurable and 
of interest for the observers. 

The importance of developing simple effective theories, approximating the behavior of systems of interest has always been clear to physicists. In 1929, just after the development of quantum mechanics and of the Dirac equation, Dirac wrote: ``The underlying physical laws necessary for the mathematical theory of a large part of physics and the whole of chemistry are thus completely known and the difficulty is only that the exact application of these laws leads to equations {\it much too complicated to be soluble}. It therefore becomes desirable that {\it approximate practical methods} of applying quantum mechanics should be developed, which can lead to an explanation of the main features of complex atomic systems without too much computation''. 

Given their practical importance, a major challenge of mathematical physics is the rigorous derivation of effective  equations starting from fundamental theories in appropriate limiting regimes. 
This should first of all justify the use of the effective equations, which are often introduced on the basis of heuristics or phenomenological arguments. Most importantly, a rigorous derivation should clarify the limits of applicability of the effective theories and should give bounds on the error of the approximation. 

We start our investigation from many body quantum mechanics. We consider systems of $N$ identical particles which can be described by a wave function $\psi_N \in L^2 (\bR^{3N})$. We are going to assume the particles to obey bosonic statistics. We will assume, in other words, the wave function $\psi_N$ to be symmetric with respect to any permutation of the $N$ particles, i.e.
\[ \psi_N (x_{\pi 1}, \dots , x_{\pi N}) = \psi_N (x_1, \dots , x_N) \]
for all permutation $\pi \in S_N$. The wave function, normalized so that $\| \psi_N \|_2 = 1$, determines the probability density $|\psi_N (x_1, \dots , x_N)|^2$ for finding particles close to $x_1, \dots , x_N$. 
More generally, an arbitrary observable, given as a self-adjoint operator on the Hilbert space $L^2 (\bR^{3N})$, can be interpreted as a random variable with probability law determined by $\psi_N$ through the spectral theorem. 

The dynamics of many body quantum systems is governed by the Schr\"odinger equation
\begin{equation}\label{eq:schr1}  i \partial_t \psi_{N,t} = H_N \psi_{N,t} \end{equation}
for the evolution of the wave function $\psi_{N,t}$. On the r.h.s. of (\ref{eq:schr1}), $H_N$ is a self-adjoint operator on $L^2 (\bR^{3N})$, known as the Hamilton operator (or simply the Hamiltonian). We will restrict our attention to Hamilton operators with two-body interaction, having the form
\begin{equation}\label{eq:ham0} H_N = \sum_{j=1}^N \left(-\Delta_{x_j}  + V_{\text{ext}} (x_j) \right) +
 \lambda \sum_{i<j}^N V (x_i -x_j) \, . \end{equation}
The sum of the Laplace operators is the kinetic part of the Hamiltonian, generating the evolution of free particles. $V_{\text{ext}}$ is an external potential and $V$ describes the (two-body) interaction among the particles (the parameter $\lambda \in \bR$ is a coupling constant, introduced here for later convenience). The Schr\"odinger equation is a linear equation, and it always has a unique global solution in $L^2 (\bR^{3N})$, obtained by applying the one-parameter group of unitary transformations generated by $H_N$ to the initial wave function $\psi_{N,0}$, i.e. $\psi_{N,t} = e^{-i H_N t} \psi_{N,0}$. What makes the Schr\"odinger equation interesting and challenging, is the fact that the number of particles $N$ involved in the dynamics is typically very large; $N$ ranges from values of the order $10^3$ in extremely dilute samples of Bose-Einstein condensates, up to value of the order $10^{23}$ in chemistry , up to even much larger values in astrophysics. For such values of $N$, it is usually impossible to extract information from the many-body Schr\"odinger equation which go beyond its 
well-posedness. For this reason it is important to find effective equations which, on the one hand, are simple to solve, either analytically or numerically, and which, on the other hand, approximate the solution of the Schr\"odinger equation in the interesting regimes. 

A simple but non trivial regime, where it is possible to obtain an effective description of the dynamics, is the so called mean field regime, which is characterized by the fact that particles experience a large number of very weak collisions. The mean field regime is realized by the Hamiltonian (\ref{eq:ham0}), when $N \gg 1$ and $|\lambda| \ll 1$, so that $N \lambda$ remains fixed, of order one.  This last condition makes sure that the total force acting on each particle, resulting from the many weak collisions, is of order one and therefore comparable with the inertia of the particle. Hence, to study the mean field regime, we will consider the dynamics  generated by the Hamiltonian 
\begin{equation}\label{eq:ham-mf} H_N^{\text{mf}} =  \sum_{j=1}^N \left(-\Delta_{x_j}  + V_{\text{ext}} (x_j) \right) +
 \frac{1}{N} \sum_{i<j}^N V (x_i -x_j) \end{equation}
in the limit of large $N$. 

To understand the emergence of a simple effective equation approximating the many body dynamics in the mean field regime, consider the evolution of a factorized (or an approximately factorized) wave function \[ \psi_{N,0} (x_1, \dots , x_N) \simeq \prod_{j=1}^N \ph (x_j) \]
for a $\ph \in L^2 (\bR^3)$. Since particles interact, factorization cannot be preserved by the time evolution. Nevertheless, because of the mean-field nature of the interaction, one can still expect factorization to be approximately restored in the limit of large $N$. We can expect, in other words, that, in the limit of large $N$, the solution of the Schr\"odinger equation $\psi_{N,t} = e^{-i H_N^{\text{mf}} t} \psi_{N,0}$ can still be approximated, in an appropriate sense, by 
\begin{equation}\label{eq:conv} \psi_{N,t} (x_1, \dots , x_N) \simeq \prod_{j=1}^N \ph_t (x_j) \, .\end{equation}
Under this assumption, it is then simple to obtain a self-consistent equation for the evolution of $\ph_t$. In fact, (\ref{eq:conv}) implies that particles are approximately independent throughout the evolution. The law of large numbers then implies that the total potential experienced by the $j$-th particles can be approximated by the average of the potential w.r.t. the distribution $|\ph_t|^2$, i.e.
\[ \frac{1}{N} \sum_{i \not = j}^N V (x_i -x_j) \simeq \int dy \,  V (x_j - y) |\ph_t (y)|^2 = (V* |\ph_t|^2) (x_j) \, . \]
Hence, if factorization is preserved in the limit of large $N$, the one-particle orbital $\ph_t$ has to evolve according to the self-consistent nonlinear Hartree equation
\begin{equation}\label{eq:hartree1}
i\partial_t \ph_t = \left(-\Delta+ V_{\text{ext}} \right) \ph_t + \left( V * |\ph_t|^2 \right) \ph_t .\end{equation}

In order to obtain precise mathematical statements about the relation between the many body Schr\"odinger evolution and the nonlinear Hartree dynamics, we need to clarify in which sense factorization is approximately preserved in the limit of large $N$. To this end, let me introduce the notion of reduced densities. For $k=1, \dots , N$, we define the $k$-particle reduced density $\gamma^{(k)}_{N,t}$ associated with the solution $\psi_{N,t} = e^{-i H_N^{\text{mf}} t} \psi_{N,0}$ of the Schr\"odinger equation by taking the partial trace of the orthogonal projection onto $\psi_{N,t}$ over the last $(N-k)$ particles. In other words, $\gamma^{(k)}_{N,t}$ is defined as a non-negative trace class operator on $L^2 (\bR^{3k})$ with integral kernel given by
\[ \begin{split} \gamma^{(k)}_{N,t} (x_1, \dots , x_k ; x'_1, \dots , x'_k) = \; &\int dx_{k+1} \dots dx_N \; \psi_{N,t} (x_1, \dots , x_k , x_{k+1}, \dots, x_N) \\ & \hspace{2cm} \times \overline{\psi}_{N,t} (x'_1, \dots, x'_k , x_{k+1}, \dots , x_N) \, . \end{split} \]
We use here the convention $\tr \, \gamma^{(k)}_{N,t} = 1$, for all $k =1, \dots , N$. For $k < N$, it is clear that the $k$-particle reduced density $\gamma^{(k)}_{N,t}$ does not contain the full information about the system. Still, knowledge of $\gamma^{(k)}_{N,t}$ is sufficient to compute the expectation of any $k$-particle observable, i.e. of any observable acting non-trivially on at most $k$ particles. In fact, if $O^{(k)}$ is a self-adjoint operator on $L^2 (\bR^{3k})$ and if $O^{(k)} \otimes 1^{(N-k)}$ denotes the operator acting as $O^{(k)}$ on the first $k$ particles and as the identity on the last $(N-k)$ particles, we find
\[ \left\langle \psi_{N,t} , \left( O^{(k)} \otimes 1^{(N-k)} \right) \psi_{N,t} \right\rangle = \tr \; \left( O^{(k)} \otimes 1^{(N-k)} \right) \, |\psi_{N,t} \rangle \langle \psi_{N,t}| = \tr \; O^{(k)} \, \gamma^{(k)}_{N,t} .\]

It turns out that the language of the reduced densities is the correct language to describe the convergence of the many body evolution towards the Hartree dynamics. 
\begin{theorem}\label{thm:mf}
Let $V$ be regular enough (see discussion below). Let $\psi_N = \ph^{\otimes N}$, for a $\ph \in H^1 (\bR^3)$. Let $\psi_{N,t} = e^{-iH^{\text{mf}}_N t} \psi_N$ be the evolution of $\psi_N$, as generated by the mean field Hamiltonian (\ref{eq:ham-mf}). Then, for every $k\in \bN$, $t \in \bR$,
\[ \gamma^{(k)}_{N,t} \to |\ph_t \rangle \langle \ph_t |^{\otimes k} \]
as $N \to \infty$, where $\ph_t$ is the solution of the nonlinear Hartree equation (\ref{eq:hartree1}), with initial data $\ph_{t=0} = \ph$. 
\end{theorem}
The first proof of this theorem was obtained by Spohn \cite{Sp} , for bounded interactions $\| V \|_\infty < \infty$. Spohn's approach was extended to potentials with a Coulomb singularity by Erd\H os and Yau \cite{EY} . These proofs were based on the direct analysis of the time-evolution of the reduced densities. {F}rom the Schr\"odinger equation for $\psi_{N,t}$, it is simple to derive a hierarchy of $N$ equations, commonly known as the BBGKY hierarchy, describing the evolution of the reduced densities:
\begin{equation}\label{eq:BBGKY} \begin{split} i\partial_t \gamma^{(k)}_{N,t} = \; &\sum_{j=1}^k \left[ -\Delta + V_{\text{ext}} (x) , \gamma^{(k)}_{N,t} \right]  + \frac{1}{N} \sum_{i<j}^k \left[ V (x_i -x_j) , \gamma^{(k)}_{N,t} \right] \\ &+ \frac{N-k}{N} \sum_{j=1}^k \tr_{k+1} \, \left[ V (x_j - x_{k+1}) , \gamma^{(k+1)}_{N,t} \right] \, .  \end{split} \end{equation}
As $N \to \infty$, the BBGKY hierarchy formally converges to the infinite hierarchy of equations
\begin{equation}\label{eq:inf-hier} \begin{split}  i\partial_t \gamma_{\infty,t}^{(k)} = \sum_{j=1}^k \left[  -\Delta + V_{\text{ext}} (x) , \gamma^{(k)}_{\infty,t} \right] + \sum_{j=1}^k \tr_{k+1} \, \left[ V (x_j - x_{k+1}) , \gamma^{(k+1)}_{\infty,t} \right] \end{split} \end{equation}
for all $k \in \bN$. It is then simple to check that this infinite hierarchy of equations has a factorized solution $\gamma^{(k)}_{\infty,t} = |\ph_t \rangle \langle \ph_t|^{\otimes k}$, given by products of solutions of the Hartree equation (\ref{eq:hartree1}). This observation suggests a strategy to obtain a proof of Theorem~\ref{thm:mf}. First, one has to show the compactness of the sequence of families $\{ \gamma_{N,t}^{(k)} \}_{k=1}^N$ with respect to an appropriate topology. Secondly, one characterizes the limit points $\{ \gamma^{(k)}_{\infty,t} \}_{k \geq 1}$ of the sequence of reduced densities as solutions of the infinite hierarchy~(\ref{eq:inf-hier}). Finally, one shows the uniqueness of the solution of the infinite hierarchy. 
This technique is very powerful, but it has few downsides. In particular, it does not give much information about the rate of the convergence towards the Hartree dynamics; this is an important point, if one is interested in applications to real physical systems, where the number of particles $N$ is very large, but of course finite. Without a bound on the rate of the convergence, it is impossible to know whether $N$ is large enough for the Hartree equation to be a good approximation.  

Motivated by these considerations, in a joint work with I. Rodnianski \cite{RS} , we gave a new proof of the convergence towards the Hartree dynamics for potentials with Coulomb singularities. Our proof was based on ideas introduced in a work of Hepp \cite{H} and then extended by Ginibre-Velo \cite{GV} . It had the advantage, compared with previous proofs, of giving precise bounds on the rate of the convergence (the technique was generalized in a joint work with L. Chen and J.-O. Lee \cite{CLS} , to obtain optimal bounds). 

More recently, Knowles and Pickl \cite{KP} developed a different approach which also covered more singular potentials. Fr\"ohlich, Knowles and Schwarz \cite{FKS} proposed an alternative point of view, expressing the convergence towards the Hartree dynamics as a Egorov-type theorem. Grillakis, Machedon and Margetis \cite{GMM1, GMM2} obtained a stronger convergence by considering a more general approximation of the full evolution in the mean field limit.

\section{The coherent states approach to mean field evolution}
\label{sec:coh}

In this section, we are going to briefly review the approach developed in collaboration with I. Rodnianski \cite{RS} . The starting point of the analysis are ideas introduced by Hepp \cite{H} to study the classical limit of quantum mechanics, which are based on the analysis of the evolution of coherent states in the Fock space.

The bosonic Fock space $\cF$ over $L^2 (\bR^3)$ is defined as the direct sum \[ \cF = \bigoplus_{n \geq 1} L^2_s (\bR^{3n}) \] where $L^2_s (\bR^{3n})$ denotes the subspace of $L^2 (\bR^{3n})$ consisting of all functions symmetric with respect to permutations of the $n$ particles. Vectors in $\cF$ are hence sequences $\{ \psi^{(n)} \}_{n \geq 1}$, with $\psi^{(n)} \in L^2_s (\bR^{3n})$. On $\cF$ we can describe states where the number of particles is not fixed. We introduce the number of particles operator $\cN$, which is defined by 
\[ (\cN\Psi)^{(n)} = n \psi^{(n)} \, \quad \text{if } \Psi = \{ \psi^{(n)} \}_{n \geq 0} \, .  \]
Eigenvectors of $\cN$ have the form $\{ 0, \dots , 0 , \psi^{(n)}, 0, \dots \}$ with only one non-vanishing entry; these vectors describe states with fixed number of particles $n$. An important example of such a state is the vacuum vector $\Omega = \{ 1, 0 , \dots \}$, describing a state with no particles at all. 

It is very useful to introduce creation and annihilation operators. For $f \in L^2 (\bR^3)$, we define the creation operator $a^* (f)$ and the annihilation operator $a(f)$ by 
\[ \begin{split} (a^* (f) \Psi)^{(n)}  (x_1, \dots , x_n) &= \frac{1}{\sqrt{n}} \sum_{j=1}^n  f(x_j) \psi^{(n-1)} (x_1, \dots , x_{j-1}, x_{j+1}, \dots , x_n) \, ,  \\ (a (f) \Psi)^{(n)} (x_1, \dots , x_n) &= \sqrt{n+1} \int dx \, f(x) \,  \psi^{(n+1)} (x, x_1, \dots, x_n) \,  \end{split} \]
for $\Psi = \{ \psi^{(n)} \}_{n \geq 0}\in \cF$. It is easy to check that $a(f)$ is the adjoint of $a^*(f)$. Creation and annihilation operators satisfy canonical commutation relations
\[ [a(f), a^* (g)] = \langle f,g \rangle_{L^2}
\quad [ a(f), a(g)] = [a^* (f) , a^* (g)] = 0 \,. \]
We also introduce operator valued distributions $a_x^*, a_x$, defined so that 
\[ a^* (f) = \int dx \, f(x) \, a_x^* \quad \text{ and } \quad a(f) = \int dx \, \overline{f} (x) \, a_x \]
for all $f \in L^2 (\bR^3)$. Expressed in terms of these operator valued distributions, the number of particles operator is given by
\[ \cN = \int dx \, a_x^* a_x \, . \]

It is important to observe that, although creation and annihilation operators are unbounded, they can be bounded with respect to the square root of $\cN$. More precisely, we have
\begin{equation}\label{eq:bda}
\| a(f) \Psi \| \leq \int dx \, |f(x)| \, \| a_x \Psi \| \leq \| f \|_2 \left( \int dx \, \| a_x \Psi \|^2 \right)^{1/2} = \| f \|_2 \, \| \cN^{1/2} \Psi \| \end{equation}
and similarly
\begin{equation}\label{eq:bda*}
\| a^* (f) \Psi \| \leq  \| f \|_2 \, \|( \cN + 1)^{1/2} \Psi \| \, . \end{equation}
 
Next, we introduce a Hamilton operator $\cH_N$ on the Fock space $\cF$. To simplify a bit the notation we will neglect the external potential and let
\begin{equation}\label{eq:ham-Fock} \cH_N = \int dx \, \nabla_x a_x^* \nabla_x a_x + \frac{1}{2N} \int dx dy \, V(x-y) \, a_x^* a_y^* a_y a_x \, . \end{equation}
Since every term in $\cH_N$ has the same number of annihilation and creation operators, $\cH_N$ commutes with the number of particles operator $\cN$. This implies that $\cH_N$ leaves each sector with fixed number of particles invariant. It is simple to compute the action of $\cH_N$ on these sectors. If $\Psi = \{ \psi^{(n)} \}_{n \geq 0}$, we have
\[ (\cH_N \Psi)^{(n)} = \cH_N^{(n)} \psi^{(n)} \]
with 
\[ \cH_N^{(n)} = \sum_{j=1}^n -\Delta_{x_j} + \frac{1}{N} \sum_{i<j}^n V(x_i - x_j) \, .\]
In particular, we observe that, when restricted on the $N$-particle sector, the Hamiltonian $\cH_N$ coincide exactly with the mean field Hamiltonian (\ref{eq:ham-mf}) (with $V_{\text{ext}} = 0$). This implies that the evolution of a Fock space vector $\Psi = \{ 0,\dots,0 , \psi_N , 0, \dots \}$ is exactly the mean field evolution discussed in the previous section, i.e. 
\[ e^{-i \cH_N t} \{ 0, \dots , 0 , \psi_N , 0, \dots \} = \{ 0, \dots , e^{-iH^{\text{mf}}_N t} \psi_N , 0 , \dots \} \,. \] The advantage of a Fock space representation is that we can consider more general initial states. In particular, we want to study the evolution of so called coherent states. 

For a $\varphi$ in $L^2 (\bR^3)$, we define the Weyl operator \[ W(\ph) = e^{a^* (\ph) - a(\ph)} \, . \]
The coherent state with wave function $\varphi$ is then described by the Fock space vector $W(\ph) \Omega$. A simple computation shows that
\[ W(\ph) \Omega = e^{-\|
\ph \|^2/2} \left\{ 1 \, , \, \ph \, , \, \frac{\ph^{\otimes 2}}{\sqrt{2!}} \, , \, \frac{\ph^{\otimes 3}}{\sqrt{3!}} \, , \dots \right\} \, .
\]
Hence coherent states do not have a fixed number of particles. Instead they are given by linear combinations of states with all possible number of particles. Since $W^* (\ph) = W(-\ph) = W^{-1} (\ph)$, Weyl operators are unitary, and therefore coherent states are always normalized $\| W(\ph) \Omega \| =1$. An important observation is that coherent states are eigenvectors of all annihilation operators. In fact, it is simple to check that
\begin{equation}\label{eq:inter} W^* (\ph) a_x W(\ph) = a_x + \ph (x), \quad \text{ and } \quad W^* (\ph) a^*_x W(\ph) = a_x^* + \overline{\ph} (x) \end{equation}
and therefore that 
\[ W^*(\ph) a(f) W(\ph) = a(f) + \langle f, \ph \rangle, \quad \text{ and } \quad W^* (\ph) a^* (f) W(\ph) = a^* (f) + \langle \ph , f \rangle  \]
for every $f \in L^2 (\bR^3)$. Since $a(f) \Omega = 0$ for all $f \in L^2 (\bR^3)$, 
these equations imply that 
\[ a (f) W (\ph) \Omega = W(\ph) W^* (\ph) a (f) W(\ph) \Omega = W(\ph) (a (f) + \langle f , \ph \rangle) \Omega = \langle f, \ph  \rangle W(\ph) \Omega \]
confirming that $W(\ph) \Omega$ is an eigenvector of $a(f)$, with eigenvalue $\langle f, \ph \rangle$, for any $f \in L^2 (\bR^3)$. As we will see shortly, the algebraic properties of coherent states and Weyl operators substantially simplify the study of their dynamics. Using (\ref{eq:inter}), we can also compute the expectation of the number of particles in the coherent state $W(\ph) \Omega$. In fact
\begin{equation}\label{eq:NW} \langle W(\ph) \Omega, \cN W(\ph) \Omega \rangle = \int dx \langle \Omega, (a^*_x + \overline{\ph} (x)) ( a_x + \ph (x)) \Omega \rangle = \| \ph \|_2^2 \,. \end{equation}
Similar computations show that the number of particles in the coherent state $W(\ph) \Omega$ is a Poisson random variable with mean and variance $\| \ph \|^2$. 

Next, we study the time-evolution of initial coherent states, as generated by the Hamiltonian (\ref{eq:ham-Fock}). In order to recover the mean field limit discussed in the previous section, the number of particles must be related with the parameter $N$ appearing in (\ref{eq:ham-Fock}). Although we cannot ask the initial coherent state to have exactly $N$ particles, we can at least require the expectation of the number of particles to be $N$. We fix therefore a $\ph \in L^2 (\bR^3)$, with $\| \ph \|_2 =1$, and we consider the time evolution 
\[ \Psi_{N,t} = e^{-i \cH_N t} W(\sqrt{N} \ph) \Omega \]
of the initial coherent state $W(\sqrt{N} \ph) \Omega$. By (\ref{eq:NW}), the expected number of particles is given by $N \| \ph \|^2_2 = N$, as desired. In particular, we are interested in the reduced  densities associated with $\Psi_{N,t}$. To make the presentation simpler, let us focus on the one-particle reduced matrix $\Gamma^{(1)}_{N,t}$ associated with $\Psi_{N,t}$. It turns out, that the integral kernel of $\Gamma^{(1)}_{N,t}$ is given by 
\[ \Gamma^{(1)}_{N,t} (x;y) = \frac{1}{\langle \Psi_{N,t} , \cN \Psi_{N,t} \rangle} \, \langle \Psi_{N,t} , a_y^* a_x \Psi_{N,t} \rangle \,. \]
Since $\cN$ is preserved by the time evolution, we have
\[ \langle \Psi_{N,t} , \cN \Psi_{N,t} \rangle = N \]
for all $t \in \bR$. Hence
\begin{equation}\label{eq:Gamma1} \Gamma^{(1)}_{N,t} (x;y) = \frac{1}{N} \, \langle a_y e^{-i \cH_N t} W (\sqrt{N} \ph) \Omega , a_x e^{-i \cH_N t} W(\sqrt{N} \ph) \Omega \rangle\,. \end{equation} 
Because of the mean field character of the interaction, we may expect the evolution of the initial coherent state to be again approximately coherent, with an evolved wave function $\ph_t$, obtained by the solution of the Hartree equation (\ref{eq:hartree1}), i.e.
\[ e^{-i \cH_N t} W(\sqrt{N} \ph) \Omega \simeq W (\sqrt{N} \ph_t) \Omega \,. \]
If this is true, $e^{-i \cH_N t} W (\sqrt{N} \ph) \Omega$ should be approximately an eigenstate of the annihilation operators $a_x, a_y$, with eigenvalues $\sqrt{N} \ph_t (x), \sqrt{N} \ph_t (y)$. Motivated by this observation, we expand the annihilation operators $a_x, a_y$ on the r.h.s. of (\ref{eq:Gamma1}), around their mean field values. We find
\begin{equation}\label{eq:Gamma2} \begin{split} \Gamma^{(1)}_{N,t} &(x;y) \\ = \; &\ph_t (x) \overline{\ph}_t (y)  \\&+ \frac{1}{N}  \left\langle (a_y - \sqrt{N} \ph_t (y)) \, e^{-i \cH_N t} W (\sqrt{N} \ph) \Omega , (a_x -\sqrt{N} \ph_t (x)) \, e^{-i \cH_N t} W(\sqrt{N} \ph) \Omega \right\rangle \\ &+ \frac{\ph_t (x)}{\sqrt{N}} \left\langle (a_y - \sqrt{N} \ph_t (y)) \, e^{-i \cH_N t} W (\sqrt{N} \ph) \Omega , \, e^{-i \cH_N t} W(\sqrt{N} \ph) \Omega \right\rangle \\ &+ \frac{\overline{\ph}_t (y)}{\sqrt{N}} \left\langle  e^{-i \cH_N t} W (\sqrt{N} \ph) \Omega , (a_x -\sqrt{N} \ph_t (x)) \, e^{-i \cH_N t} W(\sqrt{N} \ph) \Omega \right\rangle \,.\end{split} \end{equation}
We observe now that the fluctuations $(a_x - \sqrt{N} \ph_t (x))$ and $(a_y - \sqrt{N} \ph_t (y))$ can be obtained by conjugating $a_x$ and $a_y$ with evolved Weyl operators, i.e.
\[ (a_x - \sqrt{N} \ph_t (x)) = W(\sqrt{N} \ph_t) a_x W^* (\sqrt{N} \ph_t) , \]
and similarly for the fluctuations of $a_y$. We define the fluctuation dynamics as the two-parameter group of unitary transformations
\[ \cU_N (t;s) = W^*(\sqrt{N} \ph_t) e^{-i \cH_N (t-s)} W(\sqrt{N} \ph_s)\,. \]
Then (\ref{eq:Gamma2}) can be written in the compact form
\[ \begin{split} 
\Gamma^{(1)}_{N,t} &(x;y) - \ph_t (x) \overline{\ph}_t (y)  \\ = \; & \frac{1}{N}  \langle \Omega ,  \cU_N^* (t;0) \, a_y^* a_x \, \cU_N (t;0) \Omega \rangle \\ &+ \frac{\ph_t (x)}{\sqrt{N}} \langle  \Omega, \cU_N^* (t;0) \, a_y^* \, \cU_N (t;0) \Omega \rangle + \frac{\overline{\ph}_t (y)}{\sqrt{N}} \langle \Omega, \cU_N^* (t;0) \, a_x \, \cU_N (t;0) \Omega \rangle \,.
\end{split} \]
The term $\ph_t (x) \overline{\ph}_t (y)$ on the l.h.s. is just the integral kernel of the orthogonal projection $|\ph_t \rangle \langle \ph_t|$. Hence, to show the convergence $\Gamma^{(1)}_{N,t} \to |\ph_t \rangle \langle \ph_t|$, it is enough to bound the error terms on the r.h.s., and prove that they vanish in the limit of large $N$. To this end, we recall from (\ref{eq:bda}) and (\ref{eq:bda*}) that creation and annihilation operators can be bounded with respect to the square root of the number of particles operator $\cN$. The problem of proving the convergence $\Gamma^{(1)}_{N,t} \to |\ph_t \rangle \langle \ph_t|$, in the trace norm topology and with an explicit bound on the rate of the convergence, reduces therefore to establishing bounds on the growth of the expectation
\[ \langle \Omega, \cU_N^* (t;0) \, \cN \cU_N (t;0) \Omega \rangle \]
holding uniformly in $N$. 

To this end, we observe that the fluctuation dynamics $\cU_N (t;s)$ satisfies the Schr\"odinger equation
\[ i\partial_t \cU_N (t;s) = \cL_N (t) \, \cU_N (t;s) \]
with the initial condition $\cU_N (s;s) = 1$ and with the time-dependent generator
\begin{equation}\label{eq:cL} \begin{split} \cL_N (t) =  \; & \int \rd x \; \nabla_x a^*_x \nabla_x a_x + \int dx \, (V * |\ph_t|^2) (x) a_x^* a_x \\ &+ \int \rd x \rd y \; V (x-y) \ph_t (x) \overline{\ph}_t (y) \,
a_x^* a_y \\ &+ \frac{1}{2} \int \rd x \rd y \; V (x-y) \left(\ph_t (x) \ph_t
(y) \, a_x^* a^*_y + \overline{\ph}_t (x) \overline{\ph}_t (y) a_x
a_y \right) \\ &+\frac{1}{\sqrt{N}} \int \rd x \rd y \, V(x-y) \,
a_x^* \left( \overline{\ph}_t (y) a_y + \ph_t (y) a_y^* \right) a_x \\
&+\frac{1}{2N} \int \rd x \rd y \, V(x-y) \, a^*_x a^*_y a_y a_x \, . \end{split}\end{equation}
In contrast with the original Hamiltonian (\ref{eq:ham-Fock}), the generator $\cL_N (t)$ contains terms where the number of creation operators does not match the number of annihilation operators. These terms do not commute with the number of particles operator $\cN$. As a consequence, the expectation of $\cN$ is not preserved along the evolution $\cU_N$. This is hardly surprising, since $\cU_N$ defines a fluctuation dynamics, and fluctuations are expected to grow during the evolution. Although the expectation of the number of particles $\cN$ is not constant, under appropriate assumptions on the regularity of the potential, which are for example satisfied for the Coulomb case $V(x) = \pm 1/|x|$, it was shown in the paper with I. Rodnianski \cite{RS} that, for every $k \in \bN$, there exists $C,K > 0$ such that  
\begin{equation}\label{eq:no-bd} \left\langle \Psi , \cU_N^* (t;0) \, (\cN+1)^k \, \cU (t;0) \Psi \right\rangle \leq C e^{K |t|} \langle \Psi, (\cN+1)^{2k+2} \, \Psi \rangle \end{equation}
for all $t \in \bR, \Psi \in \cF$. The convergence towards the Hartree dynamics for the evolution of coherent states is then a simple corollary. 
\begin{theorem}[Mean field evolution of coherent states] \label{thm:mf-coh}
Suppose the potential $V$ satisfies the operator inequality  
\[ V^2 (x) \leq D (1-\Delta) \]
for a constant $D >0$. Let $\ph \in H^1 (\bR^3)$ and $\Psi_{N,t} = e^{-i\cH_N t} W(\sqrt{N} \ph) \Omega$. Let $\Gamma^{(1)}_{N,t}$ denote the reduced density matrix associated with $\Psi_{N,t}$. Then there exist $C,K > 0$ such that 
\begin{equation}\label{eq:Ga-conv} \tr \; \left| \Gamma^{(1)}_{N,t} - |\ph_t \rangle \langle \ph_t| \right| \leq \frac{C e^K |t|}{N} \end{equation}
for all $t \in \bR$. 
\end{theorem} 
{\it Remarks:}
\begin{itemize}
\item Since (\ref{eq:no-bd}) bounds the growth of every power of the number of particles operator, we also find, similarly to (\ref{eq:Ga-conv}), that for every $k \in \bN$ there exist $C,K > 0$ such that
\begin{equation}\label{eq:Gak-conv} \tr \;  \left| \Gamma^{(k)}_{N,t} - |\ph_t \rangle \langle \ph_t|^{\otimes k} \right| \leq \frac{C e^K |t|}{N} \end{equation}
for all $t \in \bR$.
\item {F}rom (\ref{eq:no-bd}), we see that the convergence (\ref{eq:Gak-conv}) can be extended to the evolution of initial data of the form $W(\sqrt{N} \ph) \Psi$, for arbitrary $\Psi \in \cF$ with $\langle \Psi , (\cN+1)^{2k+2} \Psi \rangle < \infty$. 
\item Writing 
\[ \ph^{\otimes N} = \frac{P_N W(\sqrt{N} \ph) \Omega}{\| P_N W(\sqrt{N} \ph) \Omega \|} \]
where $P_N$ is the orthogonal projection onto the $N$-particle sector, the bound (\ref{eq:no-bd}) can also be used to show the convergence towards the Hartree dynamics for factorized initial data. In this case, the analysis is a bit more complicated; optimal bounds on the rate of the convergence have been obtained in collaboration with L. Chen and J.-O. Lee \cite{CLS} . The same techniques can be applied to initial $N$-particle states of the form $P_N W(\sqrt{N} \ph) \Psi / \| P_N W(\sqrt{N} \ph) \Psi \|$, for arbitrary $\Psi \in \cF$ with $a(\ph) \Psi = 0$. 
\end{itemize}

It is worth observing that the coherent state approach discussed above not only implies the convergence towards the limiting Hartree evolution; instead it also give information about the behavior of the fluctuation dynamics in the limit of large $N$. In fact, it was already proven by Hepp \cite{H} and Ginibre-Velo \cite{GV} that, as $N \to \infty$, $\cU_N (t;s) \to \cU_\infty (t;s)$ strongly, where the limiting fluctuation dynamics $\cU_\infty (t;s)$ is defined by the Schr\"odinger equation \[ i\partial_t \cU_\infty (t;s) = \cL_\infty (t) \cU_\infty (t;s) \] with $\cU_\infty (s;s) = 1$ and with the time dependent generator 
\[ \begin{split} \cL_\infty (t) =  \; & \int \rd x \; \nabla_x a^*_x \nabla_x a_x + \int dx \, (V * |\ph_t|^2) (x) a_x^* a_x \\ &+ \int \rd x \rd y \; V (x-y) \ph_t (x) \overline{\ph}_t (y) \,
a_x^* a_y \\ &+ \frac{1}{2} \int \rd x \rd y \; V (x-y) \left(\ph_t (x) \ph_t
(y) \, a_x^* a^*_y + \overline{\ph}_t (x) \overline{\ph}_t (y) a_x
a_y \right) \, . \end{split} \]
The fact that $\cL_\infty (t)$ is quadratic in creation and annihilation operators implies that the limiting fluctuation dynamics $\cU_\infty (t;s)$ acts as a Bogoliubov transformation. For $f,g \in L^2 (\bR^3)$, we let $A(f,g) = a (f) + a^* (\overline{g})$. Then we have
 \[ A^* (f,g) = A (\cJ (f,g)), \quad \text{with } \quad \cJ = \left( \begin{array}{ll} 0 & J \\ J & 0 \end{array} \right)  \] 
where $J: L^2 (\bR^3) \to L^2 (\bR^3)$ is the antilinear map defined by $J f = \overline{f}$. The canonical commutation relations take the form
\[ \left[ A(f_1, g_1) , A^* (f_2 ,g_2) \right] = \langle (f_1 , g_1), \cS (f_2 , g_2) \rangle_{L^2 \oplus L^2} \qquad \text{with } S = \left( \begin{array}{ll} 1 & 0 \\ 0 & -1 \end{array} \right) \, .  \] 
A Bogoliubov transformation is a linear map $\theta : L^2 (\bR^3) \oplus L^2 (\bR^3) \to L^2 (\bR^3) \oplus L^2 (\bR^3)$ with the properties $\theta \cJ = \cJ \theta$ and $S = \theta^* S \theta$. These conditions imply that, with $B(f,g) := A (\theta (f,g))$, we have \[ B^* (f,g) = B (\cJ (f,g)) \quad \text{ and } \quad [ B(f_1, g_1), B^* (f_2, g_2)] = \langle (f_1, g_1) , S (f_2 , g_2) \rangle.\]  

One can show the existence of a two-parameter group of Bogoliubov transformations $\theta (t;s) : L^2 (\bR^3) \oplus L^2 (\bR^3) \to  L^2 (\bR^3) \oplus L^2 (\bR^3)$, such that 
\[ \cU_\infty^* (t;s) A(f,g) \cU_\infty (t;s) = A (\theta (t;s) (f,g)) \, . \]
These maps satisfy the evolution equation
\[ i\partial_t \theta (t;s) = \theta (t;s) \cA (t) \]
with $\theta (s;s) = 1$ and with the time-dependent generator
\[ \cA (t) =  \left( \begin{array}{ll} D_t  & -J B_t J \\ B_t  & -J D_t J \end{array} \right) \]
where the linear operators $D_t, B_t : L^2 (\bR^3) \to L^2 (\bR^3)$ are defined by   
\[ \begin{split} 
D_t f & = -\Delta f + (V * |\ph_t|^2) f + (V * \overline{\ph}_t f ) \ph_t \\
B_t f & = (V * \overline{\ph}_t f ) \overline{\ph}_t \, . \end{split} \]

Using this information about the limiting dynamics $\cU_\infty (t;s)$, it is possible to prove a central limit theorem for the quantum fluctuations around the mean field evolution. Consider a factorized $N$-particle initial data $\psi_N = \ph^{\otimes N}$, for some $\ph \in L^2 (\bR^3)$, and let $\psi_{N,t} = e^{-iH^{\text{mf}}_N t} \psi_N$ be its time evolution, as generated by the mean field Hamiltonian (\ref{eq:ham-mf}), with $V_{\text{ext}} = 0$ for simplicity. Let $O$ be a self-adjoint operator on $L^2 (\bR^3)$, and denote by $O^{(j)}$, for $j =1 , \dots , N$, the operator on $L^2 (\bR^{3N})$ acting as $O$ on the $j$-th particle and as the identity on the other $(N-1)$ particles. At
time $t=0$, the observables $O^{(1)}, \dots , O^{(N)}$ define a family of independent and identically distributed random variables. For $t \not = 0$, on the other hand, $O^{(1)}, \dots , O^{(N)}$ are no longer independent. The convergence (\ref{eq:Ga-conv}) easily implies that they still satisfy a law of large numbers. In a joint work with G. Ben Arous and K. Kirchpatrick \cite{BKS} , we proved that, for every $t \in \bR$, the random variables $O^{(k)}$, $k=1, \dots , N$, also satisfy a central limit theorem:
\[ \frac{1}{\sqrt{N}} \sum_{j=1}^r \left( O^{(j)} - \langle \ph_t , O \ph_t \rangle \right) \to \text{Gauss } (0, \sigma_t^2) \]
in distribution, as $N \to \infty$. So, the fluctuations around the mean field evolution are Gaussian, with a variance
\[ \sigma_t^2  = \, \left[ \left\langle \Theta_t \left(J\ph_t , \overline{J \ph_t} \right) , \Theta_t \left(J \ph_t , \overline{J}\ph_t \right) \right\rangle - \left| \left\langle \Theta_t \left(J \ph_t , \overline{J \ph_t} \right) , \frac{1}{\sqrt{2}} \, \left(\ph, \overline{\ph} \right) \right\rangle \right|^2 \right] \]
which can be expressed in terms of the Bogoliubov transfomrmation $\theta (t;s)$ discussed above.

\section{The Gross-Pitaevskii regime}

In this section, we are going to consider a different regime in which effective equations can be derived from many body quantum dynamics. The motivation comes here from the study of the time-evolution of initially trapped Bose Einstein condensates. In typical experimental settings, the condensate is initially trapped by strong magnetic field. After cooling the gas at very low temperatures, the traps are switched off and one observes the resulting evolution. The goal is to provide an effective description of the dynamics of the initially trapped condensate.

On the microscopic level, trapped Bose Einstein condensates can be described as systems of $N$ bosons with the Hamilton operator
\begin{equation}\label{eq:Htrap}  H^{\text{trap}}_N = \sum_{j=1}^N  \left(-\Delta_{x_j} + V_{\text{ext}} (x_j) \right) + \sum_{i<j}^N N^2 V (N (x_i -x_j)) \end{equation}
where $V_{\text{ext}}$ describe the trapping potential while the interaction potential $V$ is assumed to be short range and repulsive (meaning that $V \geq 0$). The interaction in (\ref{eq:Htrap}) scales with the number of particles, so that its scattering length is of the order $1/N$. Recall that the scattering length of the potential $V$ is defined through the solution of the zero-energy scattering equation 
\begin{equation}\label{eq:0en} \left( - \Delta + \frac{1}{2} V \right) f = 0 \end{equation}
with the boundary condition $f (x) \to 1$ as $|x| \to \infty$. Since $V$ has short range, it is clear that for large $|x|$, \[ f (x) \simeq  1- \frac{a_0}{|x|} \]
for a constant $a_0 >0$ which is defined to be the scattering length of $V$. Equivalently, 
\begin{equation}\label{eq:fV} 8 \pi a_0 = \int V(x) f(x) dx \, . \end{equation}
A simple scaling argument implies that $f(Nx)$ is the solution of the zero-energy scattering equation for the rescaled potential $N^2 V (Nx)$; in particular, this shows that the scattering length of $N^2 V(Nx)$ is given by $a_0 /N$. 

Lieb, Yngvason and Seiringer \cite{LSY} proved that the ground state energy $E_N$ of the Hamiltonian (\ref{eq:Htrap}) is such that
\[ \lim_{N\to \infty} \frac{E_N}{N} = \min_{\ph \in L^2 (\bR^3): \| \ph \|_2 = 1}  \cE_{\text{GP}} (\ph) \]
where we introduced the Gross-Pitaevskii energy functional
\begin{equation}\label{eq:GPen} \cE_{\text{GP}} (\ph) = \int dx \left[ |\nabla \ph (x)|^2 + V_{\text{ext}} (x) |\ph (x)|^2 + 4 \pi a_0 \, |\ph (x)|^4 \right] \, . \end{equation}
Hence, in the leading order, the ground state energy depends on the interaction potential only through its scattering length. Lieb and Seiringer \cite{LS} also showed that the ground state of $H_N^{\text{trap}}$ exhibits complete Bose-Einstein condensation in the minimizer $\phi_{\text{GP}}$ of the Gross-Pitaevskii energy functional (\ref{eq:GPen}). In other words, the one particle reduced density $\gamma^{(1)}_N$ associated with the ground state of (\ref{eq:Htrap}) is so that 
\[ \gamma^{(1)}_{N} \to |\phi_{\text{GP}} \rangle \langle \phi_{\text{GP}} | \] 
as $N \to \infty$ (in the trace norm topology). Hence, in the ground state of (\ref{eq:Htrap}) almost all particles, up to a fraction vanishing in the limit of large $N$, are in the one-particle state described by the minimizer $\phi_{\text{GP}}$ of (\ref{eq:GPen}). 

Suppose now that the boson gas is prepared in the ground state of $H^{\text{trap}}_{N}$ (by cooling the gas to almost zero temperature) and that afterwards the magnetic traps are switched off. It turns out that the Gross-Pitaevskii theory can also be used to describe the resulting evolution of the condensate, as generated by the translation invariant Hamiltonian 
\begin{equation}\label{eq:ham-GP} H_N = \sum_{j=1}^N -\Delta_{x_j} + \sum_{i<j}^N N^2 V (N (x_i - x_j)) \, . \end{equation}
In fact, it was proven in a series of works with L. Erd\H os and H.-T. Yau \cite{ESY1,ESY2,ESY3,ESY4} that, for any family of $N$-particle wave functions $\psi_N \in L^2_s (\bR^{3N})$, with finite energy per particle 
\[ \langle \psi_N, H_N \psi_N \rangle \leq C N \]
and exhibiting complete Bose-Einstein condensation 
\[ \gamma^{(1)}_N \to |\ph \rangle \langle \ph| \]
for a $\ph \in L^2 (\bR^3)$, the evolved $N$-particle wave function $\psi_{N,t} = e^{-i H_N t} \psi_N$ still exhibits complete condensation, in the sense that 
\begin{equation}\label{eq:conv-GP} \gamma^{(1)}_{N,t} \to |\ph_t \rangle \langle \ph_t | \end{equation}
as $N \to \infty$, where the dynamics of the condensate wave function $\ph_t$ is determined by the solution of the time-dependent Gross-Pitaevskii equation
\begin{equation}\label{eq:GP} i\partial_t \ph_t = -\Delta \ph_t + 8 \pi a_0 |\ph_t|^2 \ph_t \end{equation}
with initial data $\ph_{t=0} = \ph$. Another approach to show (\ref{eq:conv-GP}) has been recently proposed by Pickl \cite{P} .

The Hamiltonian (\ref{eq:ham-GP}) can be written as a mean field Hamiltonian
\[ H_N = \sum_{j=1}^N -\Delta_{x_j} + \frac{1}{N} \sum_{i<j}^N N^3 V (N (x_i -x_j)) \]
with an interaction $N^3 V (N x)$. As $N \to \infty$, at least formally,
\[ N^3 V (N x) \to b_0 \delta (x) \qquad \text{where } \quad b_0 = \int dx V(x) \, .  \]
Inserting the limiting potential $b_0 \delta (x)$ into the nonlinear Hartree equation approximating the many body dynamics in the mean field limit, we find an equation similar to (\ref{eq:GP}), but with a different coupling constant in front of the nonlinearity ($b_0$ instead of $8 \pi a_0$). The reason for the failure of the mean field analogy is that the two limits, the mean field regime on the one hand, and the Gross-Pitaevskii regime on the other hand, are very different from the physical point of view. In the mean field regime, there is a large number of very weak interactions among the particles. In the regime described by $H_N$, on the other hand, collisions are rare (particles only interact when they are very close to each others, at distances of the order $1/N$) and, at the same time, very strong. As a result, the many body evolution generated by $H_N$ produces a singular correlation structure varying on the same 
$1/N$ length scale characterizing the interaction potential. Correlations, which are negligible in the mean field limit, play here a crucial role; in particular, they are responsible for the emergence of the scattering length in the Gross-Pitaevskii equation (\ref{eq:GP}). 

To better explain this point, let us consider the first equation in the BBGKY hierarchy for the evolution of the one-particle reduced density associated with the solution of the Schr\"odinger equation $\psi_{N,t}$. Expressed in terms of the integral kernels of $\gamma^{(1)}_{N,t}$ and $\gamma^{(2)}_{N,t}$, this equation takes the form
\begin{equation}\label{eq:BBG1}
\begin{split}  i \partial_t & \gamma^{(1)}_{N,t} (x;x') \\ = &\left( -\Delta_x + \Delta_{x'} \right) \gamma^{(1)}_{N,t} (x;x') \\ &+ (N-1) \int dx_2 \left( N^2 V (N (x-x_2)) - N^2 V (N (x'-x_2)) \right) \gamma^{(2)}_{N,t} (x,x_2 ; x' , x_2) \, . \end{split}  \end{equation}
If we assume that condensation is preserved by the time evolution, the kernels of the reduced densities $\gamma^{(1)}_{N,t}$ and $\gamma^{(2)}_{N,t}$ should be approximately factorized. However, for large but finite $N$, the kernel of $\gamma^{(2)}_{N,t}$ should also contain short scale correlations between particles one and two. Assuming that these correlations can be described by the solution of the zero energy scattering equation, we obtain the ansatz
\[ \begin{split} \gamma^{(1)}_{N,t} (x_1;x'_1) &= \ph_t (x_1) \overline{\ph}_t (x'_1) \, ,  \\
\gamma^{(2)}_{N,t} (x_1,x_2 ; x'_1 , x'_2) &= f (N (x_1 - x_2)) f (N (x'_1 - x'_2)) \, \ph_t (x_1) \ph_t (x_2) \overline{\ph}_t (x'_1) \overline{\ph}_t (x'_2) \, . \end{split} \]
Inserting this ansatz in (\ref{eq:BBG1}), we obtain a self consistent equation for $\ph_t$, given by
\[i\partial_t \ph_t = -\Delta \ph_t + \left( (N-1) N^2 V (N.) f (N.) * |\ph_t|^2 \right) \ph_t \, . \]
As $N \to \infty$, (\ref{eq:fV}) implies that $(N-1)N^2 V(N.) f(N.) \to 8\pi a_0 \delta$, and we indeed obtain the correct Gross-Pitaevskii equation (\ref{eq:GP}). The presence of the correlation structure makes the proof of the convergence (\ref{eq:conv-GP}) substantially more difficult, compared with the mean field regime. Although the original strategy proposed by Spohn \cite{Sp} could still be applied, the arguments had to be modified in several points. 

The difficulty of the analysis developed in the papers with L. Erd\H os and H.-T. Yau \cite{ESY1,ESY2, ESY3}~, together with the fact that it did not provide any control on the rate of the convergence, motivated us to look for a different approach and, in particular, to understand whether the coherent states approach discussed in Section \ref{sec:coh} could also be applied in the Gross-Pitaevskii regime.  Following naively the arguments presented in Section \ref{sec:coh}, we define the Fock space Hamiltonian
\[ \cH_N = \int dx \nabla_x a_x^* \nabla_x a_x + \frac{1}{2} \int dx dy \, N^2 V (N (x-y)) \, a_x^* a_y^* a_y a_x \, . \]
We would like to compare the evolution generated by $\cH_N$ with the Gross-Pitaevskii dynamics (\ref{eq:GP}). For technical reasons, it is more convenient to compare it with the modified Gross-Pitaevskii dynamics governed by the equation
\begin{equation}\label{eq:GP-mod} i\partial_t \wt{\ph}_t = -\Delta \wt{\ph}_t + \left( N^3 V (N.) f (N.) * |\wt{\ph}_t|^2 \right) \wt{\ph}_t  \, .\end{equation}
As $N \to \infty$, it is then simple to show the convergence of the solution of (\ref{eq:GP-mod}) 
towards the solution of the original Gross-Pitaevskii equation (\ref{eq:GP}), with an error of the order $N^{-1}$ for every fixed time. 

Proceeding as in the mean field case, the comparison of the many body evolution with (\ref{eq:GP-mod}) leads to the fluctuation dynamics
\[ \cU_N (t;s) = W^* (\sqrt{N} \wt{\ph}_t) e^{-i \cH_N (t-s)} W(\sqrt{N} \wt{\ph}_s) \, . \]
As explained in Section \ref{sec:coh}, the problem of showing the convergence towards the effective evolution reduces to the problem of controlling the growth of the number of particles with respect to the evolution $\cU_N$. Computing the generator of $\cU_N$, we find the two terms
\[  \cL_N (t) =  \;  \left[ i \partial_t W^* (\sqrt{N} \ph_t) \right] W (\sqrt{N} \ph) + W^* (\sqrt{N} \ph_t) \cH_N W (\sqrt{N} \ph_t) \, . \]
In (\ref{eq:cL}), the linear (in the creation and annihilation operators) terms arising from \[ W^* (\sqrt{N} \ph_t) \cH_N W (\sqrt{N} \ph_t)\]  cancelled exactly with $\left[ i \partial_t W^* (\sqrt{N} \ph_t) \right] W (\sqrt{N} \ph)$. Because of the factor $f(N.)$ in the nonlinearity of (\ref{eq:GP-mod}), the cancellation is now not complete. We find, with $\omega = 1-f$, 
\begin{equation}\label{eq:cL-old} \begin{split} 
\cL_N (t) = & \; \sqrt{N} \int dx \Big( a_x^*  \left[ N^3 \omega (N.) V(N.) * |\ph_t|^2 \right] (x) \ph_t (x)  \\ 
& \hspace{3cm} + a_x  \left[ N^3 \omega (N.) V(N.) * |\ph_t|^2 \right] (x) \overline{\ph}_t (x) \Big) \\
&+ \text{higher order terms in $a, a^*$} .
\end{split} \end{equation}
Hence, the generator $\cL_N (t)$ contains a large term, of order $\sqrt{N}$, which does not commute with the number of particles operator $\cN$. Because of this contribution, it is impossible to prove bounds  similar to (\ref{eq:no-bd}) on the growth of the expectation of the number of particles holding uniformly in $N$. 

The reason for the failure of this naive approach is that we are effectively trying to approximate the many body evolution through an evolved coherent state, where correlations are completely absent. In other words, we are trying to bound fluctuations around the wrong correlation-free state. To control the fluctuations, we have to improve our ansatz for the many body evolution, taking in particular into account the correlation structure produced by the many body evolution. 

To this end, we introduce the integral kernel 
\[ k_t (x,y) =  - N \omega (N (x-y)) \, \ph_t (x) \, \ph_t (y) \]  
where $\omega = 1 - f$, and $f$ is the solution of the zero energy scattering equation (\ref{eq:0en}). 
For $|x-y| \gg 1/N$, we have $N \omega (N (x-y)) \simeq a_0/ |x-y|$ and therefore
\[ k_t (x,y) \simeq -\frac{a_0}{|x-y|} \ph_t (x) \ph_t (y) \, . \]
Note that $k_t$ is the kernel of a Hilbert-Schmidt operator. Using $k_t$, we define  
\[ T(t) = e^{\int dx dy \, \left( k_t (x,y) a_x^* a_y^* - \overline{k}_t (x,y) a_x a_y \right)} \, . \]
The unitary time-dependent operator $T(t)$ acts on creation and annihilation operators as a Bogoliubov transformation. In fact, we find 
\[ T(t) a(f) T^* (t) = a (\cosh_{k_t} (f)) + a^* (\sinh_{k_t} (\overline{f})) \]
where $\cosh_{k_t}, \sinh_{k_t} : L^2 (\bR^3) \to L^2 (\bR^3)$ are bounded operators, defined by the absolute convergent series 
\[ \begin{split} \cosh_{k_t} &= \sum_{j=1}^\infty \frac{k_t \overline{k}_t)^n}{(2n)!} \qquad \text{and } \quad
\sinh_{k_t} = k_t + \sum_{j=1}^\infty \frac{(k_t \overline{k}_t)^n k_t}{(2n+1)!} \, . \end{split} \]
The singularity of $k_t (x,y)$ at $x \simeq y$ disappears when we consider the powers $(k_t \overline{k}_t)^n$, $n \geq 1$. For this reason, for most purposes, we can approximate 
$\cosh_{k_t} (f) \simeq f$, $\sinh_{k_t} (f) \simeq k_t (f)$, and therefore
\begin{equation}\label{eq:TaT-ap} T^* (t) a(f) T(t) \simeq a(f) + a^* (k_t \overline{f}), \qquad \text{and } \quad T^* (t) a^* (f) T(t) \simeq a^* (f) + a (k_t \overline{f}) \, . \end{equation}

We use now the unitary operator $T(t)$ to implement the correct correlation structure, improving our approximation for the many body evolution. We introduce the new fluctuation dynamics
\[ \wt{\cU}_N (t;s) = T^* (t) W^* (\sqrt{N} \ph_t) e^{-i\cH_N (t-s)} \, W(\sqrt{N} \ph_s) T (s) \, .  \]
Then $\wt{\cU}_N (t;s)$ satisfies the Schr\"odinger equation 
\[ i\partial_t \, \wt{\cU}_N (t;s) = \wt{\cL}_N (t) \, \wt{\cU}_N (t;s) \]
with a new time-dependent generator
\begin{equation}\label{eq:cL-new} \begin{split} \wt{\cL}_N (t) = \; &\left[ i \partial_t T^* (t) \right] T(t) + T^* (t) \left[ i\partial_t W^* (\sqrt{N} \ph_t) \right] W(\sqrt{N} \ph_t) T (t) \\ &+ T^* (t) W^* (\sqrt{N} \ph_t) \cH_N W (\sqrt{N} \ph_t) T(t) \, . \end{split} \end{equation}
The derivative of the Bogoliubov transformation $T(t)$ is harmless (because it only acts on the 
solution $\wt{\ph}_t$ of the modified Gross-Pitaevskii equation, and not on the singular correlation function $\omega (N (x-y))$). Let us focus on the last two terms on the r.h.s. of (\ref{eq:cL-new}). As in (\ref{eq:cL-old}), we find a contribution linear in the creation and annihilation operators, which is now conjugated with the unitary operator $T(t)$:  
\begin{equation}\label{eq:Nlin} \sqrt{N} \int dx \, T^* (t) a_x T(t) \left[ N^3 \omega (N .) V (N.) *|\ph_t|^2 \right] (x) \, \ph_t (x) + \text{hermitian conjugate.} \end{equation}
{F}rom the last term on the r.h.s. of (\ref{eq:cL-new}), we also find a contribution cubic in creation and annihilation operators. It is given by
\[ \begin{split} 
\frac{1}{\sqrt{N}}   \int &dx dy  \; N^3 V (N(x-y)) \ph_t (y) \; T^* (t) a_x^* a_y^* a_x T(t)  + \text{h.c.} 
\\ = & \frac{1}{\sqrt{N}}   \int dx dy  \; N^3 V (N(x-y)) \ph_t (y) \; T^* (t) a_x^* a_y^* T(t) T^* (t)  a_x T(t)  + \text{h.c.} 
\\ = &  \frac{1}{\sqrt{N}}   \int dx dy  \; N^3 V (N(x-y)) \ph_t (y) \\ &\hspace{.5cm} \times 
\left(a_x^*  + \int dz \, \overline{k}_t (x, z) a_{z} \right) \left(a_y^* + \int dw \, \overline{k}_t (y,w) a_{w}\right) T^* (t)  a_x T(t)  + \text{h.c.} 
\end{split} \]
where we used the approximation (\ref{eq:TaT-ap}). Some of the terms arising from the conjugation with $T(t)$ are not normally ordered (they have a creation operator on the right of an annihilation operator). To control these contributions, we have to bring them back to normal order. Consider for example the term 
\[ \begin{split}  \frac{1}{\sqrt{N}}  \int dx dy dz \, &N^3 V (N(x-y)) \ph_t (y)  \, \overline{k}_t (x, z) a_{z}  a_y^*  T(t)^* a_x  T(t) + \text{h.c.} \\ = \, & \frac{1}{\sqrt{N}}    \int dx dy dz \, N^3 V (N(x-y)) \ph_t (y)  \, \overline{k}_t (x, z) a_y^* a_{z}   T(t)^* a_x  T(t) + \text{h.c.} \\ &+  \frac{1}{\sqrt{N}}    \int dx dy \, N^3 V (N(x-y)) \ph_t (y)  \, \overline{k}_t (x, y)  T(t)^* a_x  T(t) + \text{h.c.}
\end{split} \]
where we used the canonical commutation relation $[a_z, a_y^* ] = \delta (z-y)$. Hence, normal ordering the cubic terms generates the new linear term
\[  \begin{split} \frac{1}{\sqrt{N}}    \int dx dy \, &N^3 V (N(x-y)) \ph_t (y)  \, \overline{k}_t (x, y)  T(t)^* a_x  T(t) + \text{h.c.} \\ &= \sqrt{N} \int dx  T(t)^* a_x  T(t)  \, \left[ N^3 V (N.) \omega (N.) * |\ph_t|^2 \right] (x) \overline{\ph}_t (x) + \text{h.c.} \end{split} \]
which cancels exactly the large linear contribution (\ref{eq:Nlin}). More cancellations 
involve the quadratic and the non-normally ordered quartic terms. Using all these cancellations, and controlling the other terms appearing in $\cL_N (t)$, we obtain uniform control on the growth of the expectation of the number of particles operator with respect to the new fluctuation dynamic $\cU_N$.
As a consequence, we obtain the following theorem, establishing the convergence of the many body evolution towards the Gross-Pitaevskii equation, with a bound on the rate of the convergence.
\begin{theorem}\label{thm:GP}
Consider the time evolution 
\[ \psi_{N,t} = e^{-i \cH_N t} W (\sqrt{N} \ph) T (0) \psi \]
for a $\psi \in \cF$ with 
\[ \left\langle \psi , \left( \cN + \frac{1}{N} \, \cN^2 + \cH_N \right)  \psi \right\rangle \leq C \, .  \]
Then there exist $C,c_1, c_2 >0$ with 
\begin{equation}\label{eq:GP-thm} \tr \left| \Gamma^{(1)}_{N,t} - |\ph_t \rangle \langle \ph_t| \right| \leq \frac{C}{\sqrt{N}} \exp (c_1 \exp (c_2 |t|)) \end{equation}
for every $t \in \bR$. \end{theorem}
{\it Remarks.} 
\begin{itemize}
\item {F}rom the convergence of the one-particle reduced density $\Gamma^{(1)}_{N,t}$ towards a rank-one projection, one also obtains convergence of the higher order reduced densities, with rate $N^{-1/4}$. For every $k \geq 1$, there exist $C,c_1,c_2 > 0$ with 
\[ \tr \, \left| \Gamma^{(k)}_{N,t} - |\ph_t \rangle \langle \ph_t|^{\otimes k} \right| \leq C N^{-1/4} \exp (c_1 \exp (c_2 |t|))\, .  \]
\item Theorem \ref{thm:GP} continues to hold if the Hamiltonian contains an arbitrary external potential $V_{\text{ext}}$ (assuming $-\Delta + V_{\text{ext}}$ to be self-adjoint). 
\item The bound (\ref{eq:GP-thm}) deteriorates fast in time because it depends on high Sobolev norms of the solution $\ph_t$ of the Gross-Pitaevskii equation (\ref{eq:GP}). Assuming these norms to remain bounded in time, the estimate (\ref{eq:GP-thm}) would only deteriorate exponentially in time.
\end{itemize}

\bibliographystyle{ws-procs975x65}
\bibliography{ws-pro-sample}

\end{document}